\documentclass[twocolumn]{IEEEtran}

\usepackage{epsf}
\usepackage{graphicx}
\DeclareGraphicsExtensions{.png}
\usepackage{caption}
\usepackage{subcaption}
\usepackage{xcolor}
\usepackage{amsmath}
\usepackage{amsfonts}
\usepackage{amssymb}
\usepackage{cuted}
\usepackage{float}
\usepackage[utf8]{inputenc}
\usepackage[OT2, T1]{fontenc}
\usepackage[french,english]{babel}

\newcommand{{\calB}}{{\cal B}}

\newcommand{{\calC}}{{\cal C}}

\newcommand{{\calD}}{{\cal D}}

\newcommand{{\calE}}{{\cal E}}

\newcommand{{\calH}}{{\cal H}}

\newcommand{{\calJ}}{{\cal J}}

\newcommand{{\calM}}{{\cal M}}

\newcommand{{\calQ}}{{\cal Q}}

\newcommand{{\calP}}{{\cal P}}

\newcommand{{\calS}}{{\cal S}}

\newcommand{{\calU}}{{\cal U}}

\newcommand{{\calW}}{{\cal W}}

\newcommand{\bea}{\begin{eqnarray}}

\newcommand{\eea}{\end{eqnarray}}

\newcommand{\rmj}{{\rm j}}

\newcommand{\rmth}{{\rm th}}

\begin{document}
%
\title{Excess and Deficiency of Extreme Multidimensional Random Fields}

\author{\IEEEauthorblockN{Luk R. Arnaut} 
}

\maketitle

\begin{abstract}
Probability distributions and densities are derived for the excess and deficiency of the intensity or instantaneous energy (quasi-static power) associated with a $p$-dimensional random vector field. Explicit expressions for the exact distributions are obtained for arbitrary threshold levels, together with simple approximate functions for relatively high or low thresholds. It is shown that precise expressions only require an expansion of order $p-1$ in the ratio of the excess height to the threshold level. Numerical simulations validate the analytical results.
\end{abstract}


\section{Introduction}
In \cite[Sec. III.B]{arnaexcess}, the exceedance (height) 
of local maxima of excursions above a high threshold level $u_{\rmth}$ was statistically characterized for the intensity (squared magnitude) of a one-dimensional (1-D) Cartesian random electromagnetic (EM) field.
Results for such a single-axis component relate to measurements using linear dipole antennas. Unintentional receptors may also be susceptible to planar (2-D) or full-vectorial (3-D) EM fields.
In this paper, the 1-D result is extended to an arbitrary number of spatial dimensions $p$ \cite{arnathresh}. The general distribution is then focused on vector electric or magnetic fields with $\chi^2_n$-distributed intensities, typically at locations on ($p=1$ or $2$) or far from ($p=3$) a perfectly conducting boundary inside a reverberation chamber. 

As in \cite{arnaexcess}, all intensities will be normalized by the Cartesian mean value $\langle U_\alpha \rangle = 2\sigma^2_{X_\alpha}$ with $\alpha=x,y$ or $z$, where $\sigma_{X_\alpha}$ is the standard deviation of the in-phase or quadrature component of the Cartesian circular complex electric or magnetic field
$X_\alpha = X^\prime_\alpha +\rmj X^{\prime\prime}_\alpha$. 
For notational simplicity, such normalized quantities are denoted with a prime, i.e.,
\begin{align}
u^{\prime}_{(\rmth)} \stackrel{\Delta}{=} \frac{u_{(\rmth)}}{2\sigma^2_{X_\alpha}},~~~
h^{\pm^\prime} \stackrel{\Delta}{=} \frac{h^\pm}{2\sigma^2_{X_\alpha}}
\end{align}
(cf. sec. \ref{sec:excess} and \ref{sec:deficiency} for definitions of other symbols).
Unlike in \cite{arnaexcess}, sample values are here considered regardless of their mutual correlation. Consequently, the height $h^\pm$ is that for any sample point, not just the maximum height of an excursion.
Random variables are denoted by uppercase letters and their associated values by corresponding lowercase characters.

\section{Excess Intensity Above Threshold\label{sec:excess}}
Consider the height (positive excess) $H^+$ of an exceedance of the intensity $U$ above a fixed threshold level $u_{\rmth}$, i.e., $H^{+} \stackrel{\Delta}{=}{U}-u_{\rmth}\geq 0$. For $u_{\rmth} \gg \sigma_U$, such exceedances are relevant, e.g., to peak-level immunity testing and surges. The cumulative distribution function (CDF) of $H^{+^\prime}$ can be expressed as 
\begin{align}
F_{H^{+^\prime}}(h^{+^\prime}) &= \frac{F_{U^\prime}( u^\prime_{\rmth} + h^{+^\prime}) - F_{U^\prime}(u^\prime_{\rmth})}{1-F_{U^\prime}(u^\prime_{\rmth})}
\label{eq:exp_height_chisq_CDF}
\end{align}
with $F_{H^{+^\prime}}(h^{+^\prime}) = 0$ for $h^{+^\prime} = 0$ and $F_{H^{+^\prime}}(h^{+^\prime}) \rightarrow 1$ for $h^{+^\prime} \rightarrow +\infty$. 
Since $F_{H^{+^\prime}}(h^{+^\prime}) = F_{U^\prime}( u^\prime)$ when $u^\prime_{\rmth} \rightarrow 0$, (\ref{eq:exp_height_chisq_CDF}) is a double-barrier generalization of the single-barrier CDF of $U^\prime$. 
Specifically, whereas the complementary CDF (CCDF) $1-F_{U^\prime}( u^\prime_{\rmth})$ represents an overall tail probability as a single value with reference to $u^\prime_{\rmth}$, the CDF $F_{H^{+^\prime}}(h^{+^\prime})$ offers a detailed distribution of $U^\prime$ above $u^\prime_{\rmth}$ within this tail.

\begin{figure}[!ht] 
\begin{center} \begin{tabular}{cc}
\vspace{-3.7cm}
\hspace{-0.8cm}
\includegraphics[scale=0.48]{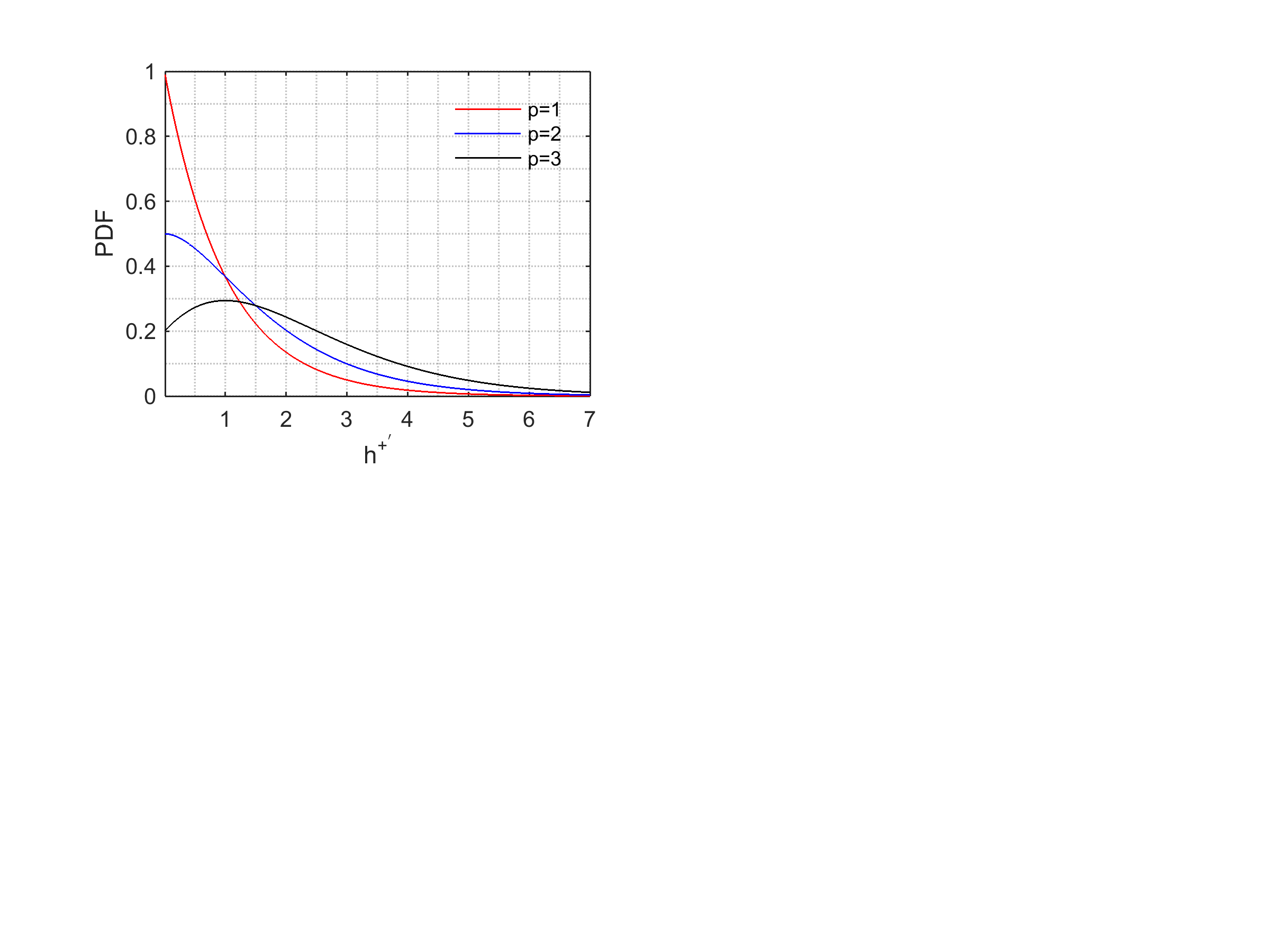}\ & \hspace{-5.7cm} \includegraphics[scale=0.48]{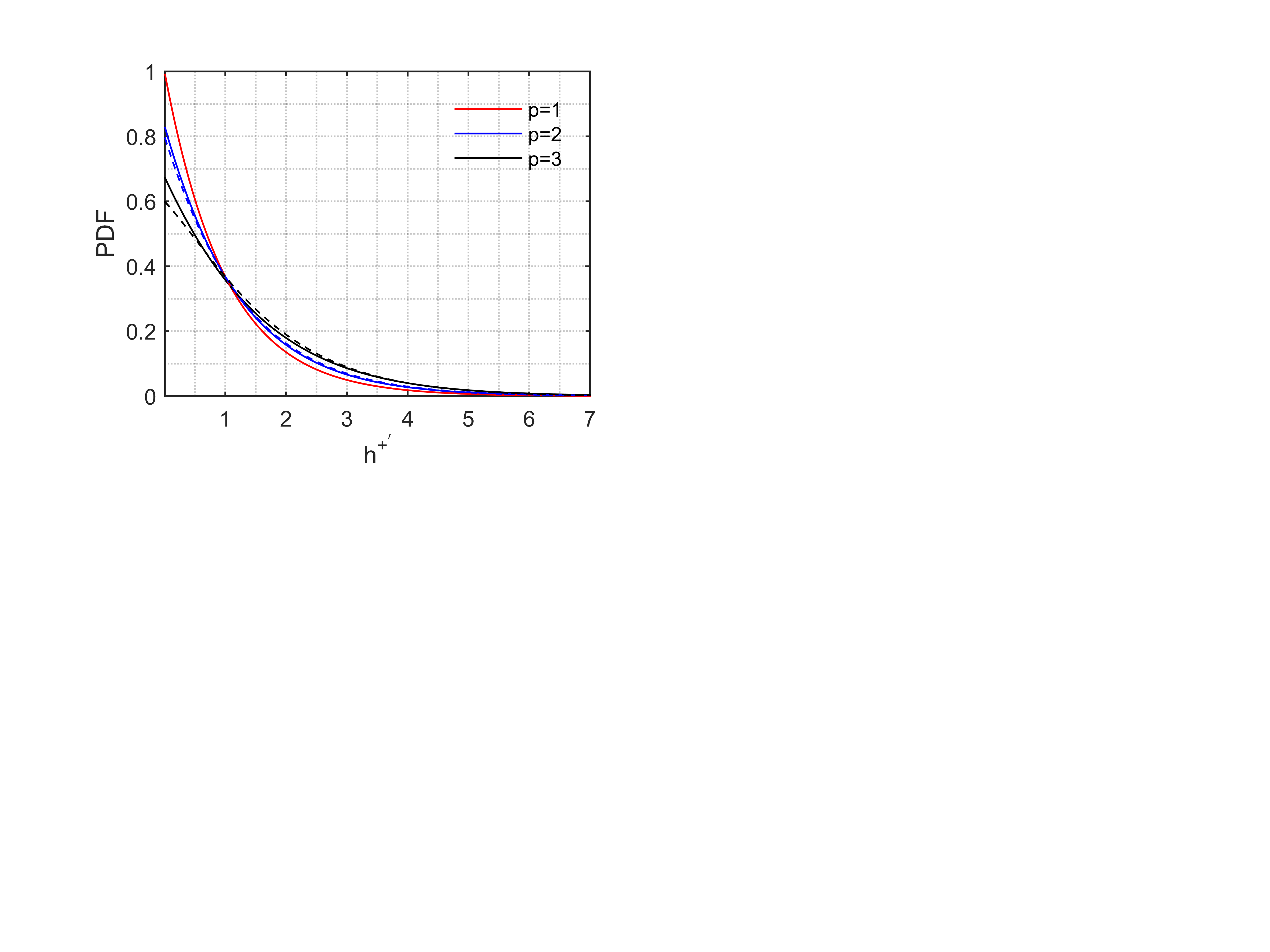}\ \\
\hspace{-5cm} (a) & \hspace{-10.5cm} (b)\\
\end{tabular} \end{center}
{
\caption{\label{fig:PDFhpls}
\small
PDFs $f_{H^{+^\prime}}(h^{+^\prime})$ for $p=1$ (red), $2$ (blue), and $3$ (black) with 
(a) $u^\prime_{\rmth} = 1$ and
(b) $u^\prime_{\rmth} = 5$.
Solid: exact PDFs (\ref{eq:fHpls_exact}); dashed: approximations (\ref{eq:fHpls_approx_largeuth}) for $u^\prime_{\rmth} \gg 1$. 
}
}
\end{figure}

For centered Gaussian $X_i$ and arbitrary $n$, the normalized intensity $U^\prime=\sum^n_{i=1} |X_i/\sigma_{X_i}|^2$
has a $\chi^2_n$ distribution, i.e., $F_{U^\prime}(u^\prime) = \gamma(n/2,u^\prime) / \Gamma(n/2)$, where $\gamma(\cdot,\cdot)$ and $\Gamma(\cdot)$ denote incomplete and complete gamma functions, respectively, and $\sigma^2_{X} = p \sigma^2_{X_\alpha}$ for a $p$-dimensional $X$. 
Upon substitution, (\ref{eq:exp_height_chisq_CDF}) can be re-expressed as an infinite series for general $n$ \cite[eq. (6.5.30)]{abra1972}. The case of $n$ odd relates to 1-D or 3-D static random fields ($n=1$ or $3$).
Here, the focus is on circular complex quasi-harmonic random fields, i.e., $n$ even ($n=2p$). 
The CDF of $U^\prime$ can then be expressed as
\begin{align}
F_{U^\prime}(u^\prime) 
&= 1 - e_{p-1} \left ( {u^\prime} \right ) \exp \left ( - {u^\prime} \right )
\end{align}
in which the truncated exponential function $e_{p-1}(\cdot)$ and its corresponding complement $e^*_{p}(\cdot)$ are defined by 
\begin{align}
\exp(x) = e_{p-1}(x) + e^*_{p}(x) 
\stackrel{\Delta}{=} 
\left ( \sum^{p-1}_{i=0} + \sum^{+\infty}_{i=p} \right ) \frac{x^i}{i!}
\label{eq:def_ep}
\end{align}
where $e_{p-1<0} (\cdot) \stackrel{\Delta}{=} 0$.
With this notation, (\ref{eq:exp_height_chisq_CDF}) becomes
\begin{align}
F_{H^{+^\prime}}(h^{+^\prime}) &=
1- \frac{e_{p-1} \left ( {u^\prime_{\rmth} + h^{+^\prime}} \right ) \exp \left ( - {h^{+^\prime}} \right )}
{e_{p-1} \left ( {u^\prime_{\rmth}} \right ) }
 \label{eq:FHpls_exact}
\end{align}
for $0 \leq h^{+\prime} < + \infty$, with probability density function (PDF)
\begin{align}
f_{H^{+^\prime}}(h^{+^\prime}) &=
\frac{\left ( u^\prime_{\rmth} + h^{+^\prime} \right )^{p-1} \exp \left ( - h^{+^\prime} \right )}
{(p-1)! ~ e_{p-1} \left ( {u^\prime_{\rmth}} \right ) }
.
\label{eq:fHpls_exact}
\end{align}
Fig. \ref{fig:PDFhpls} shows (\ref{eq:fHpls_exact}) for selected values of $p$ and $u^\prime_{\rmth}$.
For $u^\prime_{\rmth} \rightarrow +\infty$, $f_{H^{+^\prime}}(h^{+^\prime})$ converges to $\exp (- h^{+^\prime})$, irrespective of $p$.

\subsection{First-Order Expansion}
For relatively low peaks of $U$ above $u_{\rmth}$, i.e., for $h^{+^\prime} / u^\prime_{\rmth} \ll 1$ when levels near a local maximum value are of interest, a first-order approximation of (\ref{eq:FHpls_exact}) can be used, based on a binomial expansion of the argument of $e_{p-1}(u^\prime_{\rmth} + h^{+^\prime})$, viz., 
\begin{align}
e_{p-1} \left ( {u^\prime_{\rmth} + h^{+^\prime}} \right ) 
\simeq \sum^{p-1}_{i=0} \frac{\left ( {u^\prime_{\rmth} } \right )^i}{i!} \left ( 1 + i \frac{h^{+^\prime}}{u^\prime_{\rmth}} \right )
.
\label{eq:ep_firstorderapprox}
\end{align}
For $F_{H^{+^\prime}}(h^{+^\prime})$, this yields, in this approximation
\begin{align}
&~ F_{H^{+^\prime}}(h^{+^\prime} ) = 
1 -
\left [ 1 + 
\frac{e_{p-2} \left ( {u^\prime_{\rmth} } \right )}
{e_{p-1} \left ( {u^\prime_{\rmth}} \right )} {h^{+^\prime}} \right ]
 \exp \left ( - {h^{+^\prime}} \right )
 .
 \label{eq:1storderapprox_pgeneral}
\end{align}
Specifically, for 1-D, 2-D, and 3-D circular complex fields: 
\begin{itemize}
\item $p=1$: 
\begin{align}
F_{H^{+^\prime}}(h^{+^\prime}) &= 1 - \exp \left ( - {h^{+^\prime}} \right )
\label{eq:1storderexp_p1}
\end{align}
which coincides with the exact result \cite[eq. (18)]{arnaexcess} valid for arbitrary $u^\prime_{\rmth}$ and $h^{+^\prime} / u^\prime_{\rmth}$; 
\item $p=2$:
\begin{align}
F_{H^{+^\prime}}(h^{+^\prime}) &=
1 - \left ( 1 + \frac{h^{+^\prime}}{1 + {u^\prime_{\rmth}}}  \right ) \exp \left ( - {h^{+^\prime}} \right ) \label{eq:1storderexp_p2}
\\
&\simeq
1 - \left ( 1 + \frac{h^{+^\prime}}{u^\prime_{\rmth}} \right ) \exp \left ( - {h^{+^\prime}} \right )
\label{eq:1storderexp_p2_approx}
\end{align}
\item $p=3$:
\begin{align}
F_{H^{+^\prime}}(h^{+^\prime}) 
&=
1 - \left [ 1 + \frac{\left ( 1 + {u^\prime_{\rmth}} \right ) h^{+^\prime}}{1 + {u^\prime_{\rmth}} + \frac{1}{2} {u^{\prime^2}_{\rmth}} } \right ] 
\exp \left ( - {h^{+^\prime}} \right ) \label{eq:1storderexp_p3}\\
&\simeq
1 - \left ( 1 + 2 \frac{h^{+^\prime}}{u^\prime_{\rmth}} \right ) \exp \left ( - {h^{+^\prime}} \right )
\label{eq:1storderexp_p3_approx}
\end{align}
\end{itemize}
where the approximations\footnote{In this paper, all approximate PDFs and CDFs are non-normalized. In numerical evaluations, the exact expressions (\ref{eq:FHpls_exact}) and (\ref{eq:fHpls_exact}) should be used.} (\ref{eq:1storderexp_p2_approx}) and (\ref{eq:1storderexp_p3_approx}) hold for $u^\prime_{\rmth} \gg 1$.

\subsection{Second- and Higher-Order Expansions}
For exceedances that are not comparatively small, i.e., $h^{+^\prime}/u^\prime_{\rmth} \not \ll 1$, additional higher-order powers of $h^{+^\prime}/u^\prime_{\rmth}$ in the expansion of $(1+h^{+^\prime}/u^\prime_{\rmth})^i$ for $e_{p-1} [ {u^\prime_{\rmth}} ( 1 + {h^{+^\prime}}/{u^\prime_{\rmth}} ) ]$ must be retained.
Specifically, with the second-order expansion
\begin{align}
\left ( u^\prime_{\rmth} + {h^{+^\prime}} \right )^i \simeq (u^\prime_{\rmth})^i \left [ 1 + i \frac{h^{+^\prime}}{u^\prime_{\rmth}} + \frac{i(i-1)}{2!} \left ( \frac{h^{+^\prime}}{u^\prime_{\rmth}} \right )^2 \right ]
\label{eq:binomexp_2ndorder}
\end{align}
the CDF (\ref{eq:FHpls_exact}) now becomes, in this approximation
\begin{align}
&~ F_{H^{+^\prime}}(h^{+^\prime}) = \nonumber\\
&~ 
1 - \left [ 
\sum^2_{i=0} \frac{e_{p-(i+1)} (u^\prime_{\rmth})}{e_{p-1} ( u^\prime_{\rmth})} \frac{\left ( h^{+^\prime} \right )^{i}}{i!} 
\right ] \exp \left ( - {h^{+^\prime}} \right )
.
\end{align}
The CDFs for the respective 1-D, 2-D, and 3-D excess intensities are now:
\begin{itemize}
\item $p=1$: 
\begin{align}
F_{H^{+^\prime}}(h^{+^\prime}) &= 1 - \exp \left ( - {h^{+^\prime}} \right )
\label{eq:2ndorderexp_p1}
\end{align}
which is again the exact result and coincides with the first-order expansion result (\ref{eq:1storderexp_p1});
\item $p=2$:
\begin{align}
F_{H^{+^\prime}}(h^{+^\prime}) &=
1 - \left ( 1 + \frac{h^{+^\prime}}{1 + {u^\prime_{\rmth}}}  \right ) \exp \left ( - {h^{+^\prime}} \right ) \label{eq:2ndorderexp_p2}
\end{align}
which also coincides with the first-order result (\ref{eq:1storderexp_p2}); therefore,  $F_{H^{+^\prime}}(h^{+^\prime})$ for $p=2$ only requires a first-order expansion;
\item $p=3$:
\begin{align}
F_{H^{+^\prime}}(h^{+^\prime}) 
&=
1 - 
\left [ 1 + \frac{\left ( 1 + {u^\prime_{\rmth}} \right ) h^{+^\prime} + \frac{1}{2} \left ( h^{+^\prime} \right )^2}
{1 + {u^\prime_{\rmth}} + \frac{1}{2} {u^{\prime^2}_{\rmth}} } \right ] 
\nonumber\\
&~~~~~\times 
\exp \left ( - {h^{+^\prime}} \right )
\label{eq:2ndorderexp_p3}\\
&\simeq
1 - 
\left [ 1 + 2 \frac{h^{+^\prime}}{u^\prime_{\rmth}} + \left ( \frac{h^{+^\prime}}{u^\prime_{\rmth}} \right )^2
\right ] 
\exp \left ( - {h^{+^\prime}} \right )
\label{eq:2ndorderexp_p3_approx}
\end{align}
which contain additional quadratic correction terms in $( h^{+^\prime} )^2$ compared to the first-order expansion (\ref{eq:1storderexp_p3}) and its approximation (\ref{eq:1storderexp_p3_approx}), respectively.
Again, the approximation (\ref{eq:2ndorderexp_p3_approx}) holds provided $u^\prime_{\rmth} \gg 1$.
\end{itemize}

It can be easily shown that a third-order expansion of $e_{p-1} 
( {u^\prime_{\rmth}} + {h^{+^\prime}} )$ for $p=3$ results in the thus obtained $F_{H^{+^\prime}}(h^{+^\prime})$ to coincide with (\ref{eq:2ndorderexp_p3}). 
Therefore, for arbitrary $h^{+^\prime}/u^\prime_{\rmth}$, a second-order expansion is sufficient in the case of $p=3$.
Comparing (\ref{eq:2ndorderexp_p3}) with (\ref{eq:1storderexp_p3}) demonstrates that the latter expression is inaccurate, i.e., a mere first-order expansion is insufficient for this field dimensionality.
On the other hand, for $h^{+^\prime}/u^\prime_{\rmth} \ll 1$, comparing (\ref{eq:1storderexp_p3_approx}) and (\ref{eq:2ndorderexp_p3_approx}) indicates that the first-order expansion is already sufficient in $p=3$ dimensions.

\begin{figure}[!ht] 
\begin{center} \begin{tabular}{c}
\hspace{-0.8cm}
 \includegraphics[scale=0.48]{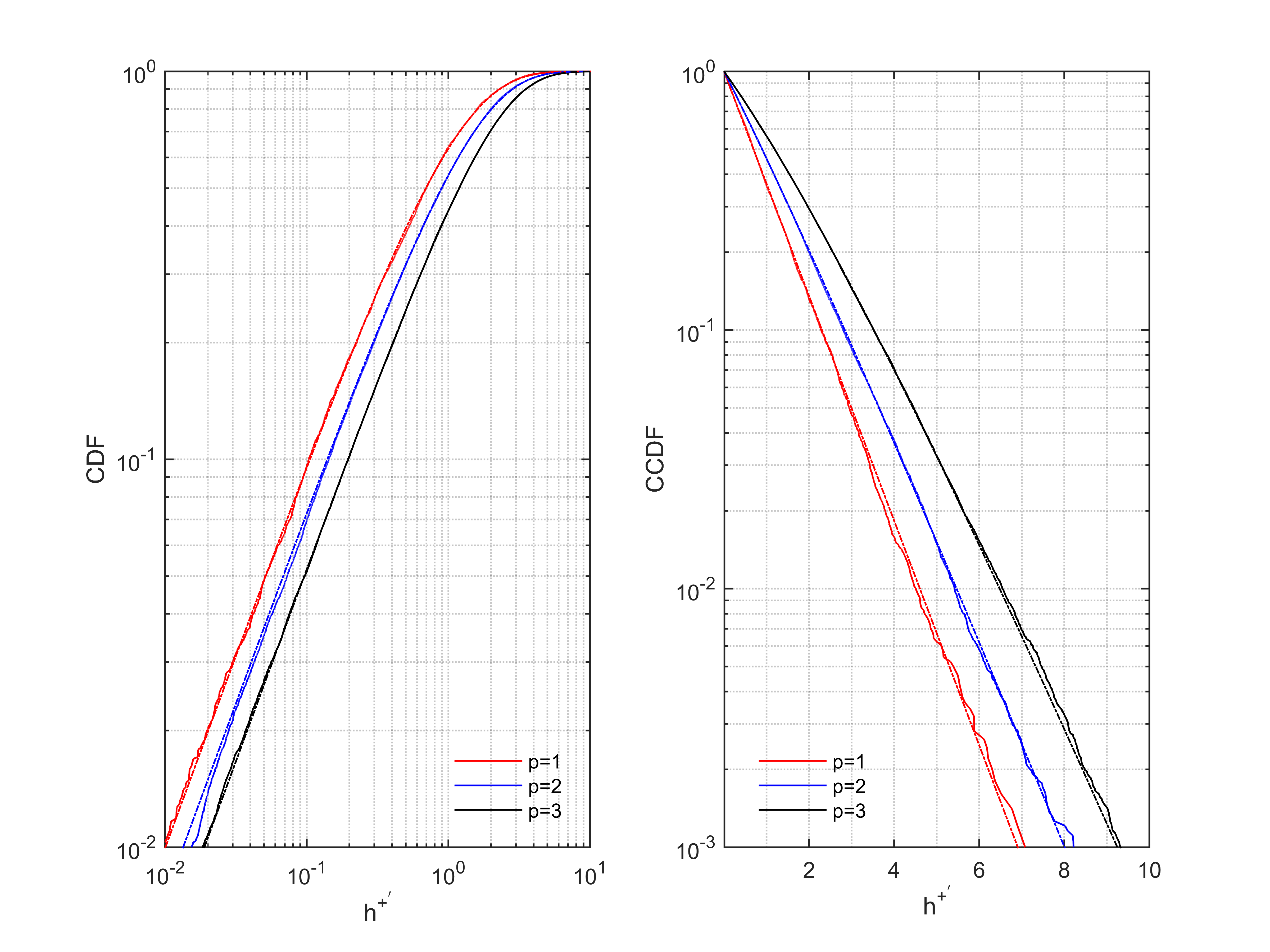}\ \\
\end{tabular} \end{center}
{
\caption{\label{fig:CDFhpls}
\small
CDFs $F_{H^{+^\prime}}(h^{+^\prime})$ (left) and 
CCDFs $1-F_{H^{+^\prime}}(h^{+^\prime})$ (right) for $p=1$ (red), $2$ (blue) and $3$ (black) with $u^\prime_{\rmth} = 3$.
Solid: MC simulation ($N=10^5$);
dashed: exact (C)CDFs (\ref{eq:FHpls_exact}). 
}
}
\end{figure}

Note that the expressions (\ref{eq:2ndorderexp_p1})--(\ref{eq:2ndorderexp_p3}) are valid for any $u^\prime_{\rmth}$, not just limited to the Poisson regime of high thresholds ($u^\prime_{\rmth}\gg 1$). In particular, $\chi^2_{2p}$ CDFs for the true (as opposed to excess) intensities $U^\prime$ \cite{kost1991} are retrieved in the limit $u^\prime_{\rmth} \rightarrow 0$, where $h^{+^\prime} \rightarrow u^\prime$. 
Fig. \ref{fig:CDFhpls} compares the exact CDFs for
$u^\prime_{\rmth} = 3$ with empirical CDFs from Monte Carlo (MC) simulations, based on $N=10^5$ 
uncorrelated 
circular Gaussian distributed samples.

For the PDF $f_{H^{+^\prime}} (h^{+^\prime})$, its right tail becomes heavier with increasing $p$ and decreasing $u^\prime_{\rmth}$. This follows from (\ref{eq:fHpls_exact}) and (\ref{eq:2ndorderexp_p1})--(\ref{eq:2ndorderexp_p3}), and is also apparent from (\ref{eq:1storderexp_p1}), (\ref{eq:1storderexp_p2_approx}) and (\ref{eq:2ndorderexp_p3_approx}) for $u^\prime_{\rmth}\gg 1$. 
From the general expression (\ref{eq:fHpls_exact}), it follows that
\begin{align}
f_{H^{+^\prime}} (h^{+^\prime}) 
&\simeq 
\left ( 1 - \frac{p-1}{u^\prime_{\rmth}} + \frac{p-1}{u^\prime_{\rmth}} h^{+^\prime}\right ) \exp \left ( - h^{+^\prime} \right )\label{eq:fHpls_approx_largeuth}\\
&\simeq 1 - \frac{p-1}{u^\prime_{\rmth}} - \left [ 1 - \frac{2(p-1)}{u^\prime_{\rmth}}\right ] h^{+^\prime}
\label{eq:fHpls_approx_largeh}
\end{align}
where (\ref{eq:fHpls_approx_largeuth}) is valid for $u^\prime_{\rmth} \gg 1$, while the linear approximation  (\ref{eq:fHpls_approx_largeh}) holds for $h^{+^\prime} \ll 1$ in addition to $u^\prime_{\rmth} \gg 1$.
Fig. \ref{fig:PDFhpls}(b) compares the approximation (\ref{eq:fHpls_approx_largeuth}) with the exact PDF (\ref{eq:fHpls_exact}) for a relatively high threshold level $u^\prime_{\rmth}=5$.

\section{Deficiency of Intensity Below Threshold \label{sec:deficiency}}
Next consider the heights of an excursion of $U$ below $u_{\rmth}$ (deficiency, negative excess), i.e., $H^{-} \stackrel{\Delta}{=}u_{\rmth} - {U} \geq 0$. This is relevant to sensitivity, susceptibility or fading testing, e.g., for detection below a noise floor at $u^\prime_{\rmth} \ll 1$. The CDF of $H^{-^\prime}$ is 
\begin{align}
F_{H^{-^\prime}}(h^{-^\prime}) &= \frac{F_{U^\prime}( u^\prime_{\rmth} ) - F_{U^\prime}(u^\prime_{\rmth} - h^{-^\prime})}{F_{U^\prime}(u^\prime_{\rmth})}
\label{eq:defic_height_chisq_CDF}
\end{align}
with $F_{H^{-^\prime}}(h^{-^\prime}) \rightarrow  0$ for $h^{-} \rightarrow 0$, i.e., $u \rightarrow u_{\rmth}$, while $F_{H^{-^\prime}}(h^{-^\prime}) \rightarrow 1$ for $h^{-} \rightarrow u_{\rmth}$, i.e., when $u^\prime \rightarrow 0$. 
Since $F_{H^{-^\prime}}( h^{-^\prime} ) = 1 - F_{U^\prime}(u^\prime)$ when $u^\prime_{\rmth} \rightarrow +\infty$, (\ref{eq:defic_height_chisq_CDF}) is a double-barrier generalization of the single-barrier CCDF of $U^\prime$.

\begin{figure}[!ht] 
\begin{center} \begin{tabular}{cc}
\vspace{-3.7cm}
\hspace{-0.8cm}
\includegraphics[scale=0.48]{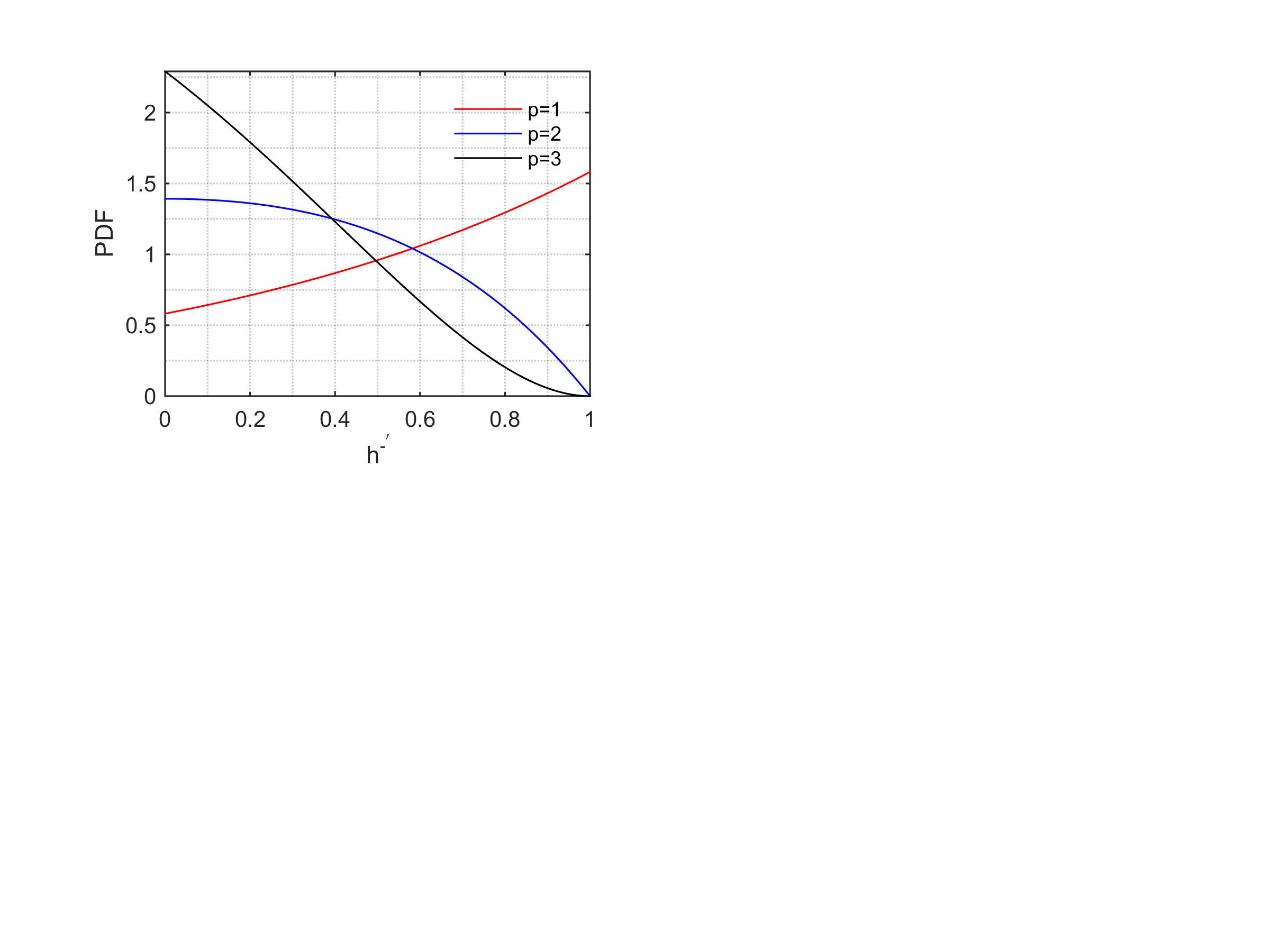}\ & \hspace{-5.7cm}
\includegraphics[scale=0.48]{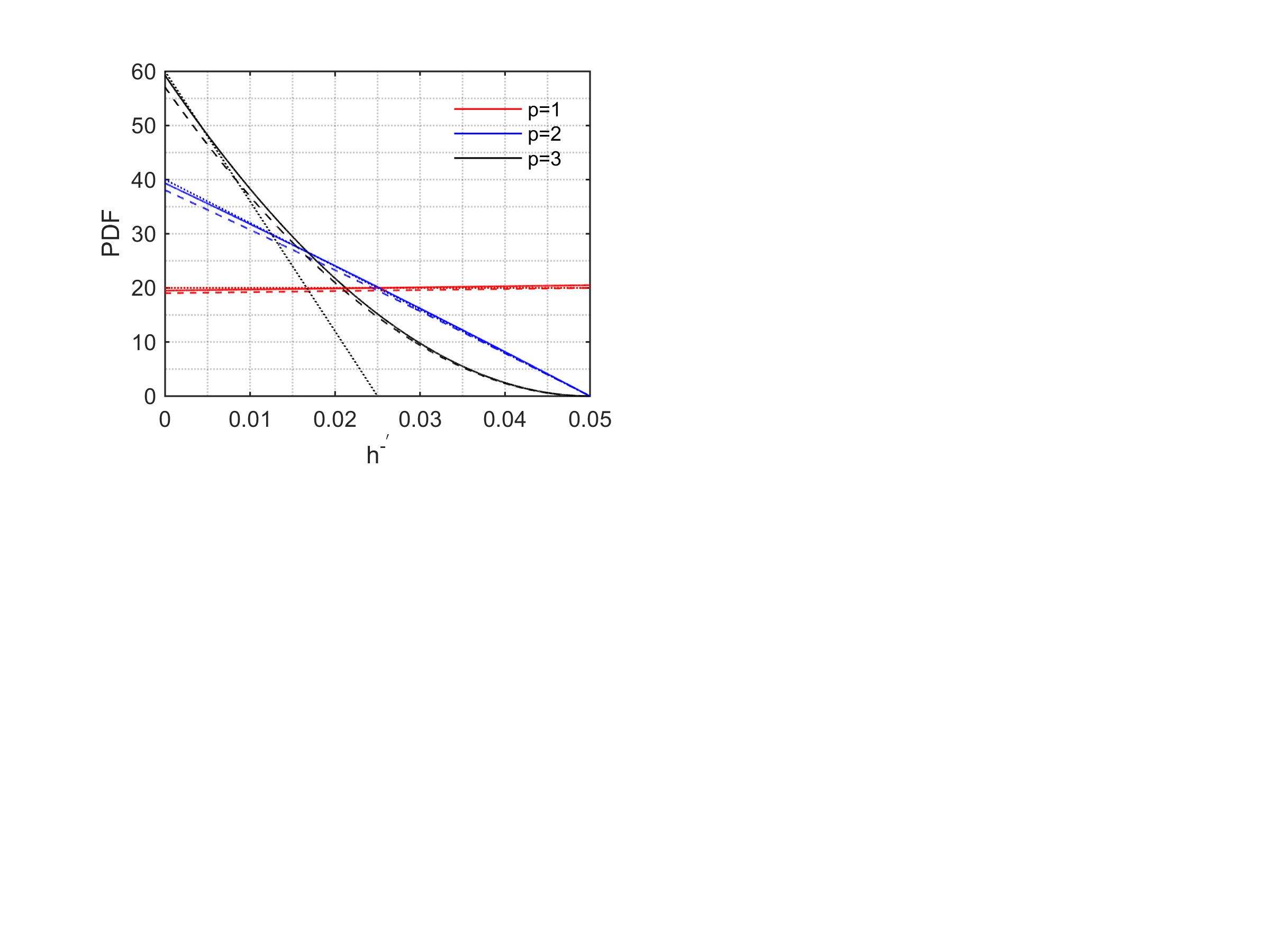}\ \\
\hspace{-5.5cm} (a) & \hspace{-10.5cm} (b)
\end{tabular} \end{center}
{
\caption{\label{fig:PDFhmin}
\small
PDFs $f_{H^{-^\prime}}(h^{-^\prime})$ for $p=1$ (red), $2$ (blue) and $3$ (black) with (a) $u^\prime_{\rmth} = 1$ and (b) $u^\prime_{\rmth} = 0.05$.
Solid: exact PDFs (\ref{eq:fHmin_exact}); 
dashed: approximations (\ref{eq:fHmin_approxsmallu}) for $u^\prime_{\rmth} \ll 1$; 
dotted: linear approximations (\ref{eq:fHmin_approxsmallhoveru}) for $u^\prime_{\rmth} \ll 1$ and $(p-1) h^{-^\prime}/u^\prime_{\rmth} \ll 1$. 
}
}
\end{figure}

For $n=2p$, (\ref{eq:defic_height_chisq_CDF}) can be expressed as
\begin{align}
&~ F_{H^{-^\prime}}(h^{-^\prime}) = \nonumber\\
&~ 1 -
\frac{1 - e_{p-1} \left ( {u^\prime_{\rmth} - h^{-^\prime}} \right ) \exp \left [ - \left ( {u^{\prime}_{\rmth}} - h^{-^\prime} \right ) \right ]}
{1 - e_{p-1} \left ( {u^\prime_{\rmth}} \right ) \exp \left ( - {u^{\prime}_{\rmth}} \right )}
 \label{eq:FHmin_exact}
\end{align}
for $0 \leq h^{-^\prime} \leq u^\prime_{\rmth}$, with corresponding PDF
\begin{align}
f_{H^{-^\prime}}(h^{-^\prime}) &=
\frac{\left ( u^\prime_{\rmth} - h^{-^\prime} \right )^{p-1} \exp \left [ - \left ( u^\prime_{\rmth} - h^{-^\prime} \right ) \right ]}
{(p-1)! ~ \left [ 1 - e_{p-1} \left ( u^\prime_{\rmth} \right ) \exp \left ( - u^\prime_{\rmth} \right ) \right ]}
\label{eq:fHmin_exact}
\end{align}
and where
\begin{align}
1 - e_{p-1} (x) \exp (- x) 
= \frac{x^p}{p!} + {\cal O} \left ( x^{p+1} \right )
\label{eq:fHmin_leadingp}
\end{align}
for $x=u^\prime_{\rmth}$ and $x=u^\prime_{\rmth} - h^{-^\prime}$ in (\ref{eq:FHmin_exact}), or  $x=u^\prime_{\rmth}$ in (\ref{eq:fHmin_exact}). Fig. \ref{fig:PDFhmin} shows (\ref{eq:fHmin_exact}) for selected values of $p$ and $u^\prime_{\rmth}$.

\begin{figure}[!ht] 
\begin{center} \begin{tabular}{c}
\hspace{-1cm}
 \includegraphics[scale=0.48]{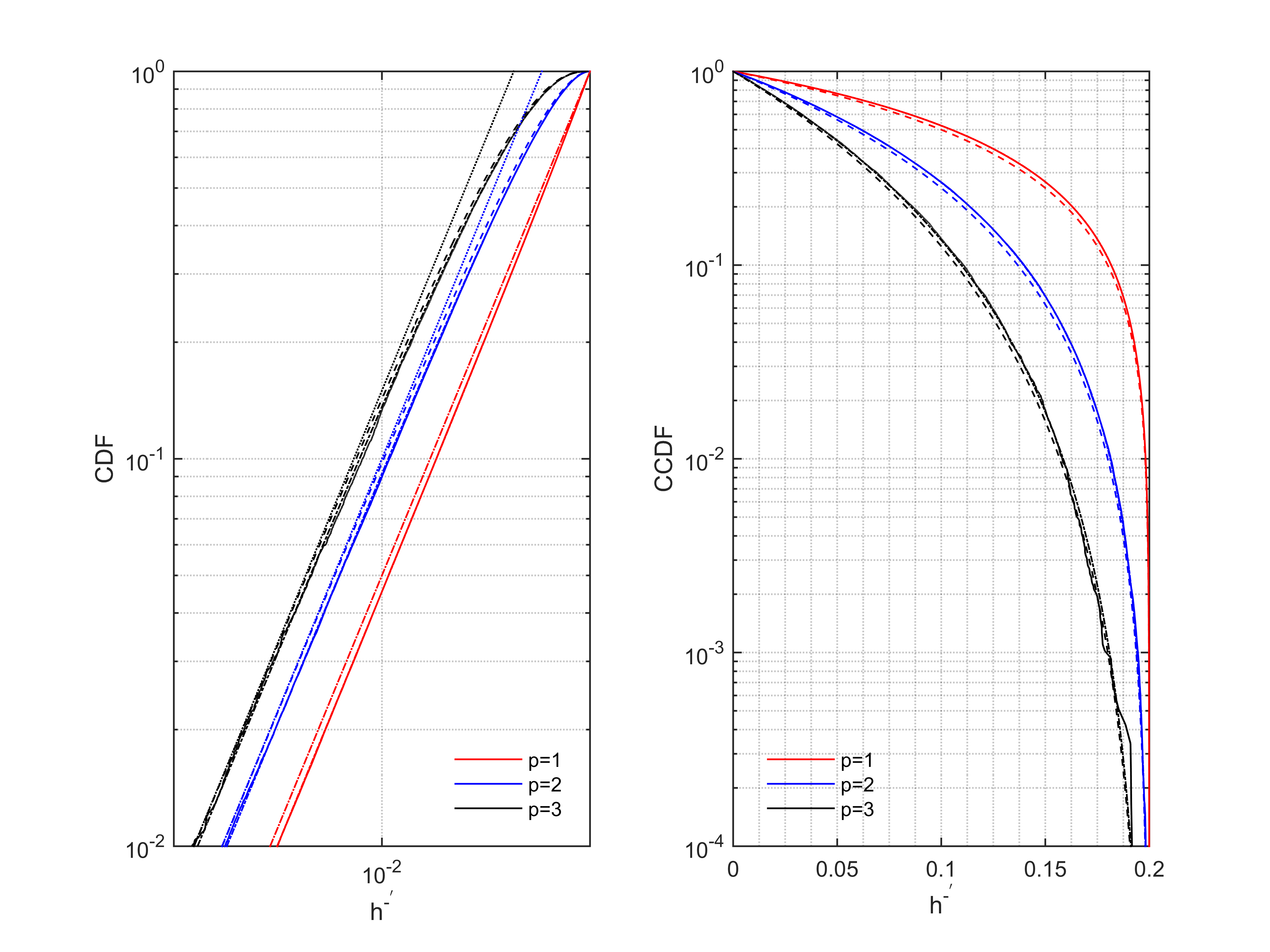}\ \\
\end{tabular} \end{center}
{\small 
\caption{\label{fig:CDFhmin}
\small
CDFs $F_{H^{-^\prime}}(h^{-^\prime})$ and CCDFs $1-F_{H^{-^\prime}}(h^{-^\prime})$ for $p=1$ (red), $2$ (blue) and $3$ (black) with
$u^\prime_{\rmth} = 0.2$.
Solid: MC simulation ($N=10^7$); 
dash-dotted: exact distribution (\ref{eq:FHmin_exact}); 
dashed: approximation (\ref{eq:FHmin_smalluth}) to order $p$ for $u^{\prime}_{\rmth} \ll 1$; 
dotted: linear approximation (\ref{eq:FHmin_smallhoveruth}) for $h^{-^\prime} / u^\prime_{\rmth} \ll 1$. 
}
}
\end{figure}

Proceeding immediately with a second-order expansion (\ref{eq:binomexp_2ndorder}) in the height-to-threshold ratio $h^{-^\prime} / u^\prime_{\rmth} < 1$, the CDFs for $p \leq 3$ are expressed as
\begin{align}
F_{H^{-^\prime}}(h^{-^\prime}) 
&= \frac{- e_{p-1} \left ( u^\prime_{\rmth} \right ) \exp \left ( - u^{\prime}_{\rmth} \right )}{1 - e_{p-1} \left ( u^\prime_{\rmth} \right ) \exp \left ( - u^{\prime}_{\rmth} \right )} \nonumber\\
&~~ + \left [ \sum^{2}_{i=0} \left (- {h^{-^\prime}}  \right )^i e_{p-(i+1)} \left ( u^\prime_{\rmth} \right ) \right ] \nonumber\\
&~~\times \frac{\exp \left ( - u^\prime_{\rmth} \right ) \sum^{\infty}_{m=0} \left ( h^{-^\prime} \right )^m / m!}{1 - e_{p-1} \left ( {u^\prime_{\rmth}} \right ) \exp \left ( - {u^{\prime}_{\rmth}} \right )}
\end{align}
whose accuracy increases as $h^{-^\prime} / u^\prime_{\rmth} \rightarrow 0$. 
Specifically, for the respective deficiencies of intensities for 1-D, 2-D, and 3-D circular complex fields: 
\begin{itemize}
\item $p=1$: 
\begin{align}
F_{H^{-^\prime}}(h^{-^\prime}) &= 1 - \frac{1 - \exp \left [ - \left ( u^\prime_{\rmth} - {h^{-^\prime}} \right ) \right ]}{1 - \exp \left ( - {u^{\prime}_{\rmth}} \right )} \nonumber\\
&\simeq 
\frac{h^{-^\prime} \exp ( - u^\prime_{\rmth} )}{1 - \exp \left ( - {u^{\prime}_{\rmth}} \right )}  
\simeq \frac{h^{-^\prime}}{u^\prime_{\rmth}}
\label{eq:2ndorderexp_p1_min}
\end{align}
\item $p=2$:
\begin{align}
F_{H^{-^\prime}}(h^{-^\prime}) 
&= 1 - \frac{1-\left ( 1 + u^\prime_{\rmth} \right ) \exp \left [ - \left ( u^\prime_{\rmth} - {h^{-^\prime}} \right ) \right ]}{1 - \left ( 1 + u^\prime_{\rmth} \right ) \exp \left ( - {u^{\prime}_{\rmth}} \right )}\nonumber\\
&~~ - \frac{h^{-^\prime} \exp \left [ - \left ( u^\prime_{\rmth} - {h^{-^\prime}} \right ) \right ]}{1 - \left ( 1 + u^\prime_{\rmth} \right ) \exp \left ( - {u^{\prime}_{\rmth}} \right )} \nonumber\\
&~ \simeq \frac{u^\prime_{\rmth} h^{-^\prime} \exp (-u^\prime_{\rmth} )}{1 - \left ( 1 + u^\prime_{\rmth} \right ) \exp \left ( - u^{\prime}_{\rmth} \right )} 
\simeq \frac{2 h^{-^\prime}}{u^\prime_{\rmth}}
\label{eq:2ndorderexp_p2_min}
\end{align}
\item $p=3$:
\begin{align}
&F_{H^{-^\prime}}(h^{-^\prime}) 
= \nonumber\\
&1 - \frac{1 - \left ( 1 + u^\prime_{\rmth} + \frac{1}{2} u^{\prime^2}_{\rmth} \right ) \exp \left [ - \left ( u^\prime_{\rmth} - {h^{-^\prime}} \right ) \right ]}{1 - \left ( 1 + u^\prime_{\rmth} + \frac{1}{2} u^{\prime^2}_{\rmth}\right ) \exp \left ( - {u^{\prime}_{\rmth}} \right )}\nonumber\\
&~~ - \frac{\left ( 1 + u^\prime_{\rmth} \right ) h^{-^\prime} \exp \left [ - \left ( u^\prime_{\rmth} - {h^{-^\prime}} \right ) \right ]}{1 - \left ( 1 + u^\prime_{\rmth} + \frac{1}{2} u^{\prime^2}_{\rmth} \right ) \exp \left ( - {u^{\prime}_{\rmth}} \right )} \nonumber\\
&~~ + \frac{\frac{1}{2} \left ( h^{-^\prime}  \right )^2 \exp \left [ - \left ( u^\prime_{\rmth} - {h^{-^\prime}} \right ) \right ]}{1 - \left ( 1 + u^\prime_{\rmth}  + \frac{1}{2} u^{\prime^2}_{\rmth} \right ) \exp \left ( - {u^{\prime}_{\rmth}} \right )} \nonumber\\
&\simeq \frac{\frac{1}{2} u^{\prime^2}_{\rmth} h^{-^\prime} \exp (-u^\prime_{\rmth} )}{1 - \left ( 1 + u^\prime_{\rmth} + \frac{1}{2} u^{\prime^2}_{\rmth} \right ) \exp \left ( - u^{\prime}_{\rmth} \right )} 
\simeq \frac{3 h^{-^\prime}}{u^\prime_{\rmth}}
\label{eq:2ndorderexp_p3_min}
\end{align}
\end{itemize}
where the penultimate approximations in (\ref{eq:2ndorderexp_p1_min})--(\ref{eq:2ndorderexp_p3_min}) assume $h^{-^\prime} \ll 1$, while the final approximations assume additionally $u^\prime_{\rmth} \ll 1$. These final approximations also follow immediately from (\ref{eq:FHmin_exact}) with (\ref{eq:fHmin_leadingp}) as 
\begin{align}
F_{H^{-^\prime}}(h^{-^\prime}) = 1 - \left ( 1-\frac{h^{-^\prime}}{u^\prime_{\rmth}} \right )^p 
+ {\cal O} \left [ \left (\frac{h^{-^\prime}}{u^\prime_{\rmth}} \right )^{p+1} \right ]
\label{eq:FHmin_smalluth}
\end{align}
for $u^\prime_{\rmth} \ll 1$ (implying $h^{-^\prime} \ll 1$), and subsequently 
\begin{align}
F_{H^{-^\prime}}(h^{-^\prime}) 
\simeq \frac{p}{u^\prime_{\rmth}} h^{-^\prime} 
\label{eq:FHmin_smallhoveruth}
\end{align} 
if additionally $h^{-^\prime}/u^\prime_{\rmth} \ll 1$. 
In Fig. \ref{fig:CDFhmin}, these expressions are compared with the exact CDF (\ref{eq:FHmin_exact}) and with MC simulation results for $u^\prime_{\rmth} = 0.2$. Residual differences remain as a result of choosing $u^\prime_{\rmth} \not \ll 1$ in this example, to demonstrate its effect.

For the PDF $f_{H^{-^\prime}}(h^{-^\prime})$, 
it follows from (\ref{eq:fHmin_exact})--(\ref{eq:fHmin_leadingp}), to leading order in $ h^{-^\prime} / u^\prime_{\rmth}$, that
\begin{align}
f_{H^{-^\prime}} (h^{-^\prime}) 
&\simeq \frac{p}{u^\prime_{\rmth}} \left ( 1 - \frac{h^{-^\prime}}{u^\prime_{\rmth}} \right )^{p-1} \exp \left [ - u^\prime_{\rmth}\left ( 1 - \frac{h^{-^\prime}}{u^\prime_{\rmth}} \right ) \right ]\label{eq:fHmin_approxsmallu}\\
&\simeq \frac{p}{u^\prime_{\rmth}} 
\left [ 1 - \frac{(p-1) h^{-^\prime}}{u^\prime_{\rmth}} \right ]\simeq \frac{p}{u^\prime_{\rmth}}
\label{eq:fHmin_approxsmallhoveru}
\end{align}
where (\ref{eq:fHmin_approxsmallu}) holds for $h^{-^\prime} \leq u^\prime_{\rmth} \ll 1$, while (\ref{eq:fHmin_approxsmallhoveru}) additionally assumes $h^{-^\prime}/u^\prime_{\rmth} \ll 1$ and the final approximation in (\ref{eq:fHmin_approxsmallhoveru}) holds provided $(p-1) h^{-^\prime}/u^\prime_{\rmth} \ll 1$. Since $0\leq h^{-^\prime} \leq u^\prime_{\rmth}$, this final approximation applies across the entire domain of $h^{-^\prime}$ only when $p=1$. 
These linear approximations are shown in Fig. \ref{fig:PDFhmin}(b).
The comparison of (\ref{eq:fHmin_approxsmallhoveru}) with (\ref{eq:fHpls_approx_largeh}) demonstrates the difference in the effect of the zero lower bound for $u^\prime_{\rmth}$ on $f_{H^{-^\prime}}$ versus the unlimited upper bound for $u^\prime_{\rmth}$ on $f_{H^{+^\prime}}$.

\section{Conclusion}
In testing for immunity, susceptibility or fading, the consideration of exceedances (i.e., exceedingly high or low values near the absolute maximum or minimum) for the field intensity or energy allows for a more accurate classification, modelling and estimation of the behaviour and distributions of extreme values. 
This situation also arises when actual values are off-scale (i.e., outside the instrumentation's measurement range) or beyond the measurement horizon (duration of the interval of observation), so that peak values need to be estimated from the available limited data.
With the aid of (\ref{eq:exp_height_chisq_CDF}) and (\ref{eq:defic_height_chisq_CDF}), the distribution of such a surplus or shortage may be determined empirically, {\it a fortiori} without prior knowledge or assumption of a theoretical distribution model such as $\chi^2_{2p}$ in this paper.

For circular Gaussian fields, the analysis shows that the CDF of the positive or negative excess intensity for a 1-D field is independent of the expansion order in the height-to-threshold ratio $h^{\pm^\prime}/u^\prime_{\rmth}$, i.e., 
(\ref{eq:2ndorderexp_p1}) and (\ref{eq:2ndorderexp_p1_min}).
For higher-dimensional vector fields ($p>1$), an expansion order not higher than $p-1$ in $h^{\pm^\prime}/u^\prime_{\rmth}$ is already sufficient to obtain a precise CDF for this and all lower field dimensionalities, independently of the value of $u^\prime_{\rmth}$ relative to $1$.
For $h^{\pm^\prime}/u^\prime_{\rmth} \ll 1$, a first-order expansion is always sufficient, irrespective of $p$.

Because of the so-called curse of dimensionality, an empirical determination of distributions of $H^{+^\prime}$ and $H^{-^\prime}$ for $p>1$ would require excessive amounts of data to achieve acceptably high definition and accuracy. This underlines the merit of the theoretical tail distributions derived and analyzed in this paper.

\end{document}